\documentclass[showpacs,twocolumn,aps,pra,longbibliography,superscriptaddress,notitlepage]{revtex4-2}

\usepackage{graphicx}
\usepackage{physics}
\usepackage{wrapfig}
\usepackage{changepage}
\usepackage{amsmath,amssymb,amsthm,mathrsfs,amsfonts,dsfont,mathtools}
\usepackage[caption=false]{subfig}
\usepackage{epsfig}
\usepackage{color,xcolor}
\usepackage{adjustbox}
\usepackage{tabularx}
\usepackage{array}
\usepackage{multirow}
\usepackage{tabu}
\usepackage{bm}
\usepackage{enumerate}
\usepackage[colorlinks=true,linkcolor=blue,citecolor=blue,urlcolor=blue,pdfauthor={ },pdftitle={ },pdfsubject={ },pdfkeywords={ }]{hyperref}
\usepackage{newtxtext,newtxmath}
\usepackage{makecell} 
\usepackage{url}
\usepackage{booktabs}
\usepackage{nicefrac} 
\usepackage{makeidx}  
\usepackage{algorithm}
\usepackage{algpseudocode}
\usepackage{cleveref}
\usepackage{mwe}

\usepackage{comment}
\usepackage{xurl}

\newcommand{\pz}{\sigma_z}

\usepackage{color}
\definecolor{dred}{rgb}{.8,0.2,.2}
\definecolor{ddred}{rgb}{.8,0.5,.5}
\definecolor{dblue}{rgb}{.2,0.2,.8}
\definecolor{dgreen}{rgb}{.2,0.5,.2}

\begin{document}
\title{Neural quantum embedding via deterministic quantum computation with one qubit}

\author{Hongfeng Liu}
\thanks{These authors contributed equally to this work.}
\affiliation{Department of Physics and Guangdong Basic Research Center of Excellence for Quantum Science, Southern University of Science and Technology, Shenzhen 518055, China}

\author{Tak Hur}
\thanks{These authors contributed equally to this work.}
\affiliation{Department of Statistics and Data Science, Yonsei University, Seoul 03722, Republic of Korea}

\author{Shitao Zhang}
\thanks{These authors contributed equally to this work.}
\affiliation{Shenzhen Institute for Quantum Science and Engineering, Southern University of Science and Technology, Shenzhen 518055, China}

\author{Liangyu Che}
\affiliation{Department of Physics and Guangdong Basic Research Center of Excellence for Quantum Science, Southern University of Science and Technology, Shenzhen 518055, China}

\author{Xinyue Long}
\affiliation{Quantum Science Center of Guangdong-HongKong-Macao Greater Bay Area, Shenzhen 518045, China}

\author{Xiangyu Wang}
\affiliation{Department of Physics and Guangdong Basic Research Center of Excellence for Quantum Science, Southern University of Science and Technology, Shenzhen 518055, China}

\author{Keyi Huang}
\affiliation{Department of Physics and Guangdong Basic Research Center of Excellence for Quantum Science, Southern University of Science and Technology, Shenzhen 518055, China}

\author{Yu-ang Fan}
\affiliation{Department of Physics and Guangdong Basic Research Center of Excellence for Quantum Science, Southern University of Science and Technology, Shenzhen 518055, China}

\author{Yuxuan Zheng}
\affiliation{Department of Physics and Guangdong Basic Research Center of Excellence for Quantum Science, Southern University of Science and Technology, Shenzhen 518055, China}

\author{Yufang Feng}
\affiliation{Department of Physics and Guangdong Basic Research Center of Excellence for Quantum Science, Southern University of Science and Technology, Shenzhen 518055, China}

\author{Xinfang Nie}
\affiliation{Department of Physics and Guangdong Basic Research Center of Excellence for Quantum Science, Southern University of Science and Technology, Shenzhen 518055, China}
\affiliation{Quantum Science Center of Guangdong-HongKong-Macao Greater Bay Area, Shenzhen 518045, China}

\author{Daniel K. Park}
\email{dkd.park@yonsei.ac.kr}
\affiliation{Department of Statistics and Data Science, Yonsei University, Seoul 03722, Republic of Korea}
\affiliation{Department of Applied Statistics, Yonsei University, Seoul 03722, Republic of Korea}

\author{Dawei Lu}
\email{ludw@sustech.edu.cn}
\affiliation{Department of Physics and Guangdong Basic Research Center of Excellence for Quantum Science, Southern University of Science and Technology, Shenzhen 518055, China}
\affiliation{Shenzhen Institute for Quantum Science and Engineering, Southern University of Science and Technology, Shenzhen 518055, China}
\affiliation{Quantum Science Center of Guangdong-HongKong-Macao Greater Bay Area, Shenzhen 518045, China}

\begin{abstract}
Quantum computing is expected to provide exponential speedup in machine learning. However, optimizing the data loading process, commonly referred to as quantum data embedding, to maximize classification performance remains a critical challenge. In this work, we propose a neural quantum embedding (NQE) technique based on deterministic quantum computation with one qubit (DQC1). Unlike the traditional embedding approach, NQE trains a neural network to maximize the trace distance between quantum states corresponding to different categories of classical data. Furthermore, training is efficiently achieved using DQC1, which is specifically designed for ensemble quantum systems, such as nuclear magnetic resonance (NMR). We validate the NQE-DQC1 protocol by encoding handwritten images into NMR quantum processors, demonstrating a significant improvement in distinguishability compared to traditional methods. Additionally, after training the NQE, we implement a parameterized quantum circuit for classification tasks, achieving 98\% classification accuracy, in contrast to the 54\% accuracy obtained using traditional embedding. Moreover, we show that the NQE-DQC1 protocol is extendable, enabling the use of the NMR system for NQE training due to its high compatibility with DQC1, while subsequent machine learning tasks can be performed on other physical platforms, such as superconducting circuits. Our work opens new avenues for utilizing ensemble quantum systems for efficient classical data embedding into quantum registers.
  
\end{abstract}

\maketitle


\textit{Introduction}---Quantum machine learning (QML) presents a compelling frontier for data science, pushing the field beyond the capabilities of classical information processing systems. While QML is inherently suited for learning from quantum data~\cite{biamonte2017quantum,QSVM,QPCA,10.1038/s43588-022-00311-3}, the majority of data analysis tasks in modern society targets classical data. Thus, developing an effective approach for applying QML to classical data is of critical importance for a wide range of real-world problems. Addressing this challenge entails the intricate task of devising optimal feature mappings of classical data to quantum states, known as quantum data embedding, tailored to the specific datasets being used. Practical QML also faces the ongoing challenge of building universal and fault-tolerant quantum hardware. Various approaches have been proposed to leverage non-fault-tolerant quantum computers in QML~\cite{Preskill2018quantumcomputingin,Havlicek2019,benedetti_parameterized_2019,cerezo2020variational,bharti2022noisy}. However, the potential contribution of subuniversal quantum computers to this field remains unexplored. Developing useful applications of these less powerful yet more feasible quantum devices for QML represents a significant milestone toward practical quantum advantage.

To address these challenges, we propose a method for optimizing quantum data embedding using deterministic quantum computation with one qubit (DQC1)~\cite{DQC1PhysRevLett.81.5672}. In particular, we focus on enhancing quantum data embedding for binary classification, a fundamental task in data analysis~\cite{PhysRevLett.114.140504,PhysRevLett.126.110502}. For an effective classifier, it is essential that training data points from different classes exhibit large distances, while points within the same class remain closely clustered in the feature space~\cite{bartlett2017spectrally,hur2023neural,hur2024understanding}. To achieve this under limited quantum resources, we introduce a classical-quantum hybrid algorithm that leverages DQC1 to train a neural network, enabling it to learn a quantum feature map that meets these criteria. 

DQC1 is a subuniversal model of quantum computation featuring a single probe qubit initialized with non-zero purity, $n$ uniformly random bits, the ability to implement arbitrary unitary transformations, and the measurement of Pauli observables on the probe qubit. Although DQC1 is limited to the single-qubit state preparation and measurement, it can solve certain problems faster than classical computers~\cite{DQC1PhysRevLett.81.5672,PhysRevA.72.042316,PhysRevLett.92.177906,DQC1complexity,PhysRevA.97.032327}. A notable example is estimating the normalized trace of an $n$-qubit unitary operator. This capability can be directly applied to efficiently estimating the Hilbert-Schmidt (HS) inner product~\cite{Nielsen:2011:QCQ:1972505} between two $n$-qubit unitary operators, denoted by $U_1$ and $U_2$, which is defined as $\langle U_1,U_2\rangle_{\mathrm{HS}} = \mathrm{Tr}(U_1^{\dagger}U_2)/(2^n)$. The core idea of our method is to optimize quantum data embedding by training a neural network with the objective of maximizing (or minimizing) the HS inner product for unitary operators that map data from the same (or different) class to the quantum feature space. We refer to this technique as neural quantum embedding via DQC1 (NQE-DQC1). Ensemble quantum information processors, such as those based on spin ensembles and magnetic resonance~\cite{doi:10.1073/pnas.94.5.1634,doi:10.1098/rsta.2011.0352,PhysRevLett.123.030502,Kyungdeock2015,Lu2016,qiu2021quantum}, are ideal platforms for implementing DQC1 protocols. Moreover, once the NQE is trained on ensemble systems, subsequent QML tasks can seamlessly transition to other physical platforms. This flexibility establishes ensemble quantum systems as a powerful resource for effective QML on classical data. 

\begin{figure}[t]
    \centering
    \includegraphics[width=0.9\columnwidth]{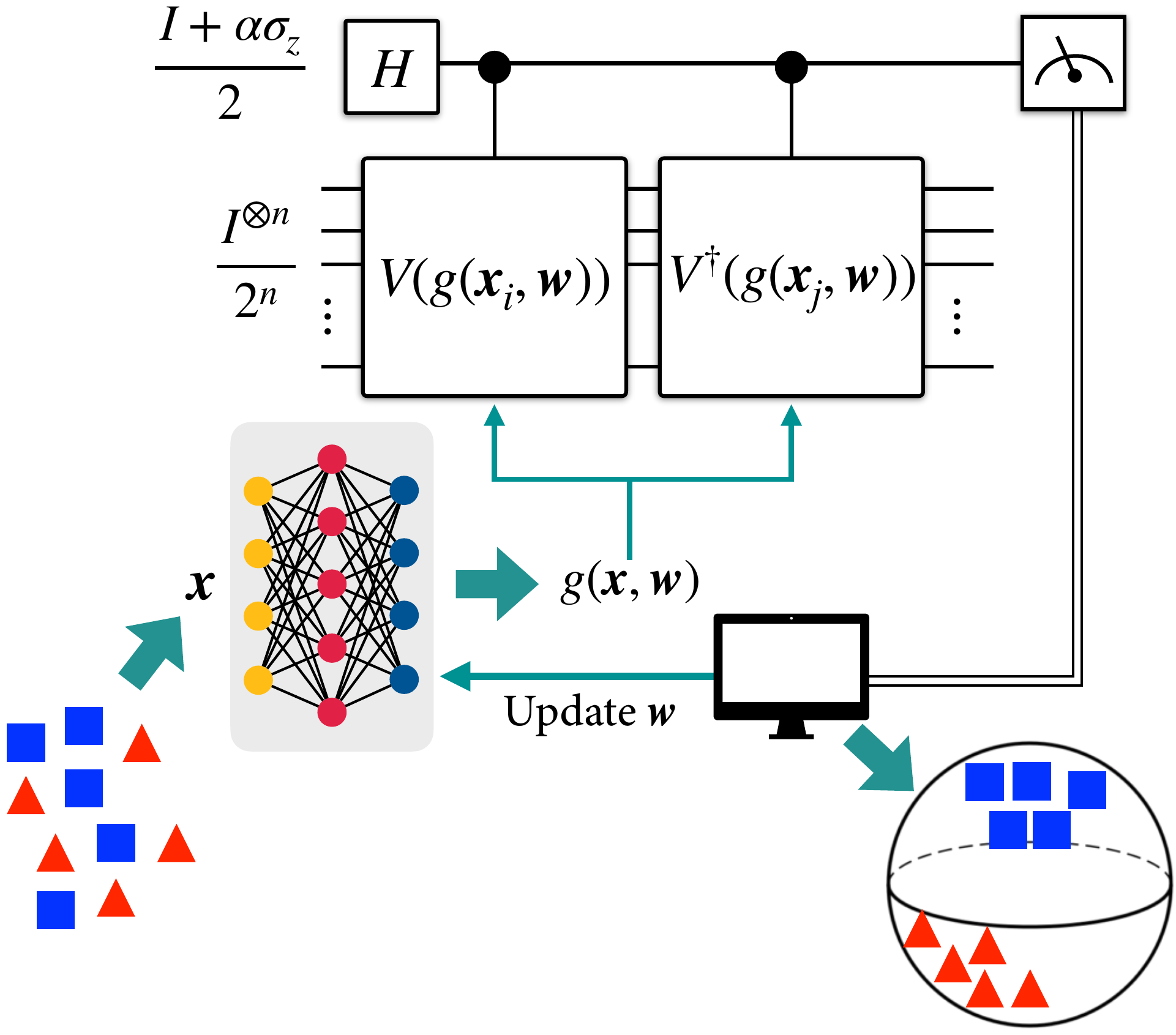}
    \caption{Schematic of the neural quantum embedding via DQC1, illustrating the process for identifying the optimal quantum feature map for classification tasks. Data points with different labels are represented as squares and triangles.}
    \label{fig:nqe-dqc1}
\end{figure}

\begin{figure}[b]
    \centering
    \includegraphics[width=1\linewidth]{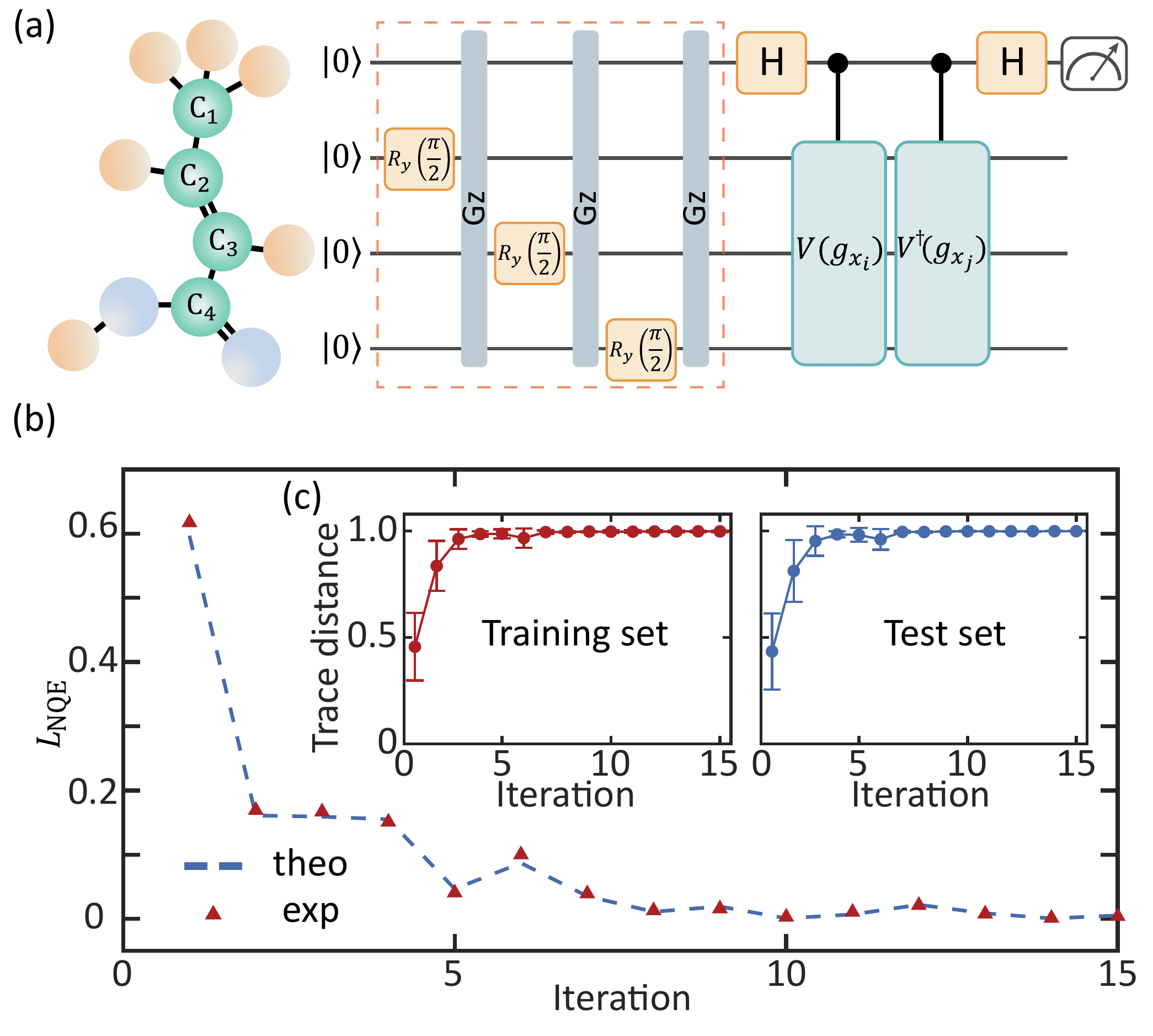}
    \caption{ (a) Experimental quantum circuit for NQE-DQC1 training. The experimental setup uses a four-qubit NMR quantum processor comprising four carbon spins. The operations within the red dashed box initialize the system to $\vert 0 \rangle \langle 0 \vert \otimes I/2^n$ with $n=3$. The gray bar represents a 1 ms $z$-gradient pulse, which is applied to eliminate coherence. The unitary operators $V$ and $V^{\dagger}$ are determined by the output of the neural network $g$, where $x_i$ serves as the image data input, and $\boldsymbol{w}$ represents the trainable parameters. $H$ denotes the Hadamard gate, and measurements are performed solely on the probe qubit. (b) Training loss histories are presented, with the solid line and diamonds representing the simulated and experimental mean values of the training loss across 10 pairs of images, respectively. (c) Trace distance is used as a metric to evaluate the distinguishability of images from different categories. Twenty pairs of images are randomly selected from both the training and test datasets. Solid dots indicate the mean trace distance, while error bars denote the standard deviation.}
    \label{nqe}
\end{figure}

\textit{Protocol overview}---Let the class label of the input data $\boldsymbol{x}_i$ be $y_i\in \lbrace +1,-1\rbrace$. The quantum data embedding circuit must be constructed to maximize the fidelity between quantum states corresponding to input data points with the same label ($y_i = y_j$) and minimize it for those with different labels. This can be formulated as maximizing the distinguishability between $\rho_{\pm} = \sum_{i}^{m_{\pm}}|\boldsymbol{x}_i\rangle\langle \boldsymbol{x_i}|/m_{\pm}$, which denotes the quantum state representation of data samples labeled $\pm 1$~\cite{Bae_2015}. Here, $m_{\pm}$ indicates the number of data points in class $\pm 1$. For quantum data embedding, this is equivalent to finding a quantum feature map that 
maximizes the trace distance, $\Vert \rho_+-\rho_-\Vert_1/2$~\cite{hur2023neural}. 

Our method utilizes a classical neural network to optimize quantum data embedding according to these criteria. The objective function subject to minimization is 
\begin{equation}
\label{eq:loss}
    L_\text{NQE}=\sum_{ij}\left\{\frac{1}{2^n}\mathrm{Tr}\left[V(g_{\boldsymbol{x}_{i}})V^\dagger(g_{\boldsymbol{x}_{j}})\right]-\frac{1+y_iy_j}{2}\right\}^2,
\end{equation}
where $g_{\boldsymbol{x}_{i}} = g(\boldsymbol{x}_{i},\boldsymbol{w})$ is defined as the output of the neural network, $\boldsymbol{w}$ is the set of trainable parameters, and $V$ is the quantum feature map that encodes $V(g_{\boldsymbol{x}})|\psi\rangle = |\boldsymbol{x}\rangle$. Then $|\boldsymbol{x}\rangle$ is provided as an input to a QML algorithm, such as the quantum neural network~\cite{benedetti_parameterized_2019,grant_hierarchical_2018,cong_quantum_2019,hur2022quantum,Park_2023,kim2023classical,PerezSalinas2020datareuploading,kim2024expressivity} or the quantum kernel method~\cite{Havlicek2019,RigorousRobustQSpeedUp,huang_power_2021}, to construct a classifier.

The first term in Eq.~(\ref{eq:loss}) is the HS inner product between two $n$-qubit unitary operators, which quantifies the similarity between two data points. Finding the solution to minimizing Eq.~(\ref{eq:loss}) corresponds to learning $\boldsymbol{w}$ such that the HS inner product is large when the two data points belong to the same class and small otherwise.  It is also related to the distance defined by the Frobenius norm, $\Vert A \Vert_F = \sqrt{\mathrm{Tr}(A^{\dagger}A)}$, since
$
    \Vert V(g_{\boldsymbol{x}_{i}})-V(g_{\boldsymbol{x}_{j}})\Vert^2_F
 = 2 - 2\mathrm{Re}\left[\mathrm{Tr}\left(V(g_{\boldsymbol{x}_{i}})V(g_{\boldsymbol{x}_{j}})^{\dagger}\right)\right].
$

Thus, the task of minimizing Eq.~\eqref{eq:loss} reduces to efficiently computing the HS inner product within the objective function. As previously noted, DQC1 provides an efficient means for estimating this metric~\cite{Nielsen:2011:QCQ:1972505}. Additionally, $V$ can be optimized by employing a classical neural network and deep learning techniques. This approach overcomes the limitation imposed by completely positive and trace-preserving (CPTP) maps, which cannot increase the trace distance between quantum states. Specifically, any subsequent CPTP map $\Lambda$ satisfies the inequality $\|\Lambda(\rho_+) - \Lambda(\rho_-)\|_1 \leq \|\rho_+ - \rho_-\|_1$~\cite{e23050625}. Further details are provided in the Supplementary Material~\cite{supp}.

The schematic of NQE-DQC1 is depicted in Fig.~\ref{fig:nqe-dqc1} using the task of embedding a subset of the MNIST database of handwritten digits~\cite{lecun2010mnist} as an example. In the following, we demonstrate a proof-of-principle implementation of NQE-DQC1 on a nuclear magnatic resonance (NMR) quantum processor.

\textit{Experimental scheme}---We experimentally demonstrate the NQE-DQC1 protocol on an NMR platform. The experiments are conducted on a Bruker 300 MHz spectrometer at room temperature. The sample used is $^{13}$C-labeled trans-crotonic acid dissolved in $d_6$-acetone, with the four $^{13}$C nuclear spins forming a four-qubit quantum processor~\cite{PhysRevLett.129.070502,PhysRevLett.132.210403,cheng2024experimental,lin2024hardware}. The molecular structure of the sample is shown in Fig.~\ref{nqe}(a).
The internal Hamiltonian governing this system is 
$\mathcal{H}_{\text{NMR}}=-\sum_{i}\omega_i\sigma^i_z/2+\sum_{i,j}\pi J_{ij}\sigma^i_z\sigma^j_z/2$, 
where $\omega_i/2\pi$ is the Larmor frequency of the $i$th spin, and $J_{ij}$ represents the scalar coupling between the $i$th and $j$th spins. The specific parameters are provided in Supplemental Material~\cite{supp}.
In the NMR experimental setup, single-qubit rotations are implemented through transverse radio-frequency pulses, while two-qubit gates are realized via free evolutions of the interactions in $\mathcal{H}_{\text{NMR}}$. The experimental control accuracy is further enhanced using gradient-based optimization to generate shaped pulses~\cite{khaneja2005optimal}.
The experimental NQE-DQC1 circuit is depicted in Fig.~\ref{nqe}(a), where C$_1$ serves as the probe qubit, and the remaining spins act as encoding qubits.
The entire NQE-DQC1 experiment consists of three stages.

(i) Initialization. 
The entire system is first prepared into $\vert 0000 \rangle$ from thermal equilibrium using the spatial averaging method~\cite{cory1998nuclear}. Subsequently, each of the encoding qubits undergoes a 2 ms $R_y(\pi/2)$ pulse ($\pi/2$ rotation about the $y$-axis) and a 1 ms $z$-gradient pulse. This operation initializes the system to $\rho_0 = \vert 0 \rangle \langle0 \vert \otimes I/2^n$, where $I$ is the $8 \times 8$ identity matrix and $n = 3$.

(ii) NQE training. 
The dataset used in this experiment comes from 500 images of the MNIST dataset~\cite{lecun2010mnist}, which depicts handwritten digits ``0'' and ``1''. To construct the training dataset, ten pairs of images are randomly selected, with each pair comprising two images that may either belong to the same category or different categories.
The gray-scale values of the images are then extracted and subjected to principal component analysis (PCA). It is important to note that PCA preprocessing is not strictly necessary, as the NQE training at this stage is able to handle the ``dimensionality reduction''~\cite{hur2023neural}. However, to reduce the number of training parameters of the classical neural network, we preprocess the images using classical PCA before implementing the quantum circuit.

After applying PCA, we obtain two $1 \times 5$ vectors, e.g., $\boldsymbol{x}_1$ and $\boldsymbol{x}_2$, corresponding to the two images in each pair. Each vector $\boldsymbol{x}$ is then non-linearly mapped to a quantum state on the encoding qubits via $\vert \boldsymbol{x} \rangle = V(g_{\boldsymbol{x}})\vert 0 \rangle$. Here, $V$ denotes a $ZZ$-feature map~\cite{Havlicek2019}, expressed as
\begin{equation}
V(\boldsymbol{\phi}) = \left\{ \text{exp}\left[i\sum_{k} \boldsymbol{\phi}_k Z_k + \boldsymbol{\phi}_{n + k} Z_k Z_{k+1}\right]H^{\otimes n}\right\}^M,
\end{equation}
where $H$ is the Hadamard gate, $Z_k$ is the Pauli-$z$ operator on the $k$th spin, $\boldsymbol{\phi}_k$ represents the $k$th entry of the vector $\boldsymbol{x}$, and $M$ is the number of layers, which we set $M = 1$. 
As required by the DQC1 protocol, the implementation of this feature map depends on the state of the probe qubit C$_1$. Starting from $\rho_0$, the following gate sequence, $\mathcal{U}=H_1\mathcal{V}^\dagger_c(g_{\boldsymbol{x}_2})\mathcal{V}_c(g_{\boldsymbol{x}_1})H_1$, is applied (right to left).
Here, the Hadamard gate $H_1$ can create superposition (or vice versa) on C$_1$, while $\mathcal{V}_c(g_{\boldsymbol{x}})=\vert0\rangle\langle0\vert\otimes I+\vert1\rangle\langle1\vert\otimes V(g_{\boldsymbol{x}})$ represents a controlled feature map conditioned on the state of C$_1$. The corresponding circuit is illustrated in Fig.~\ref{nqe}(a).

(iii) Measurement and optimization. 
Here, we only need to measure the probe qubit along its $z$-direction, as $\langle \pz \rangle = \Re{\text{Tr}[V(g_{\boldsymbol{x}_1}) V^\dagger(g_{\boldsymbol{x}_2})]}/2^n$. The experimental result is then substituted into Eq.~\eqref{eq:loss} to compute the loss value $L_\text{NQE}$.
Subsequently, the neural network $g$ is optimized using the gradient $\nabla L_\text{NQE}$, which is computed from the average loss value obtained across the ten pairs of image data. 
In the next iteration, ten new pairs of images are randomly selected, and the above steps are repeated to further optimize the neural network.

\begin{figure}[t]
    \centering
    \includegraphics[width=1\linewidth]{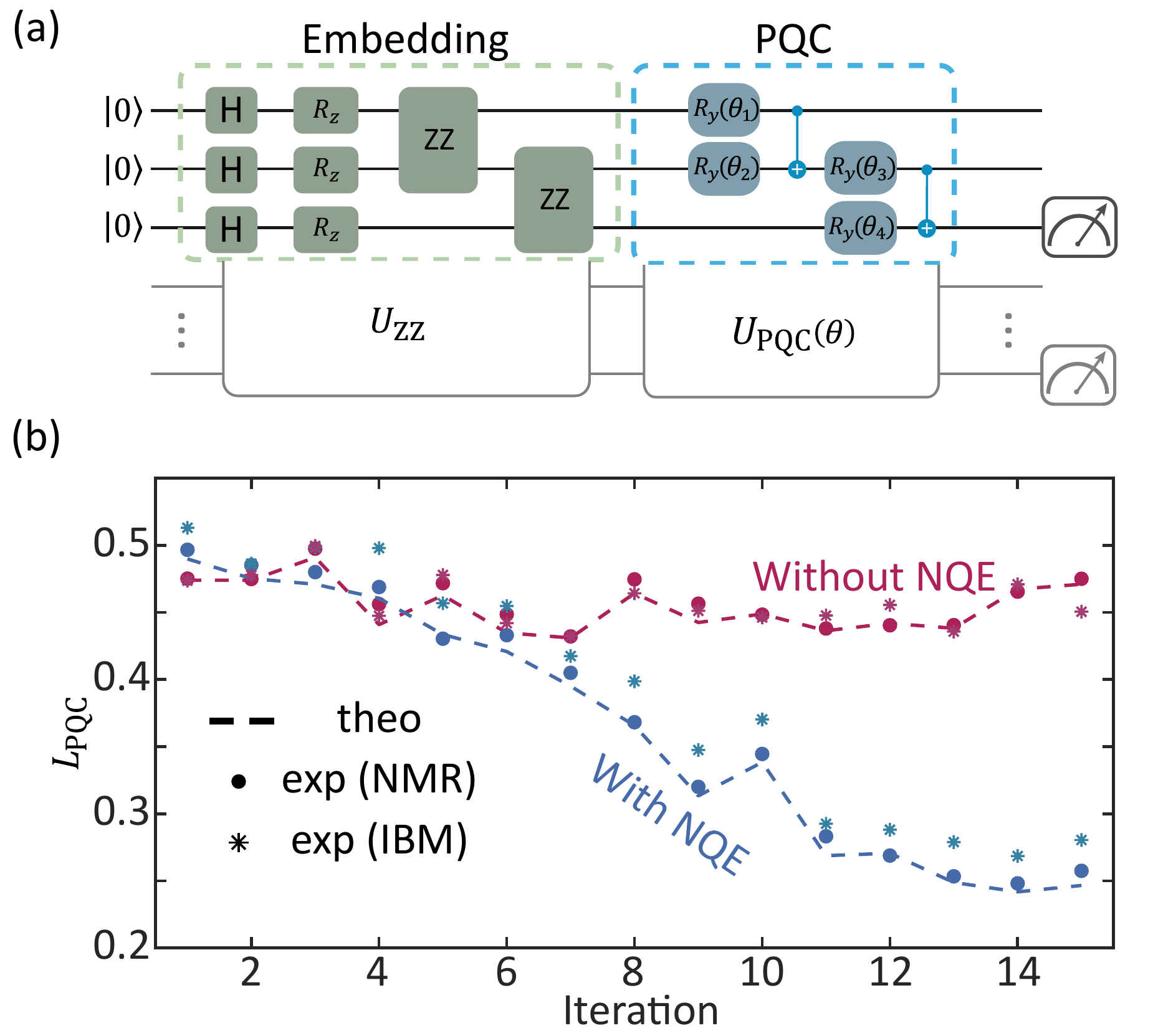}
    \caption{(a) Quantum circuit for classification tasks. Classical images are first encoded into quantum states using NQE and then processed by the PQC classifier. The PQC consists of two layers, incorporating four $\theta$ parameters and two CNOT gates. Measurements of $\langle\pz\rangle$ are performed on the third spin. (b) Optimization results of the PQC. The optimization results for the PQC are presented, where solid lines, dots, and dashed lines correspond to the loss $L_{\text{PQC}}$ from numerical simulations, NMR experiments, and IBM experiments, respectively. Red and blue denote results obtained under traditional $ZZ$-feature embedding (without NQE) and with NQE encoding, respectively.}
    \label{PQC}
\end{figure}

\begin{figure*}[t]
    \centering
    \includegraphics[width=1\linewidth]{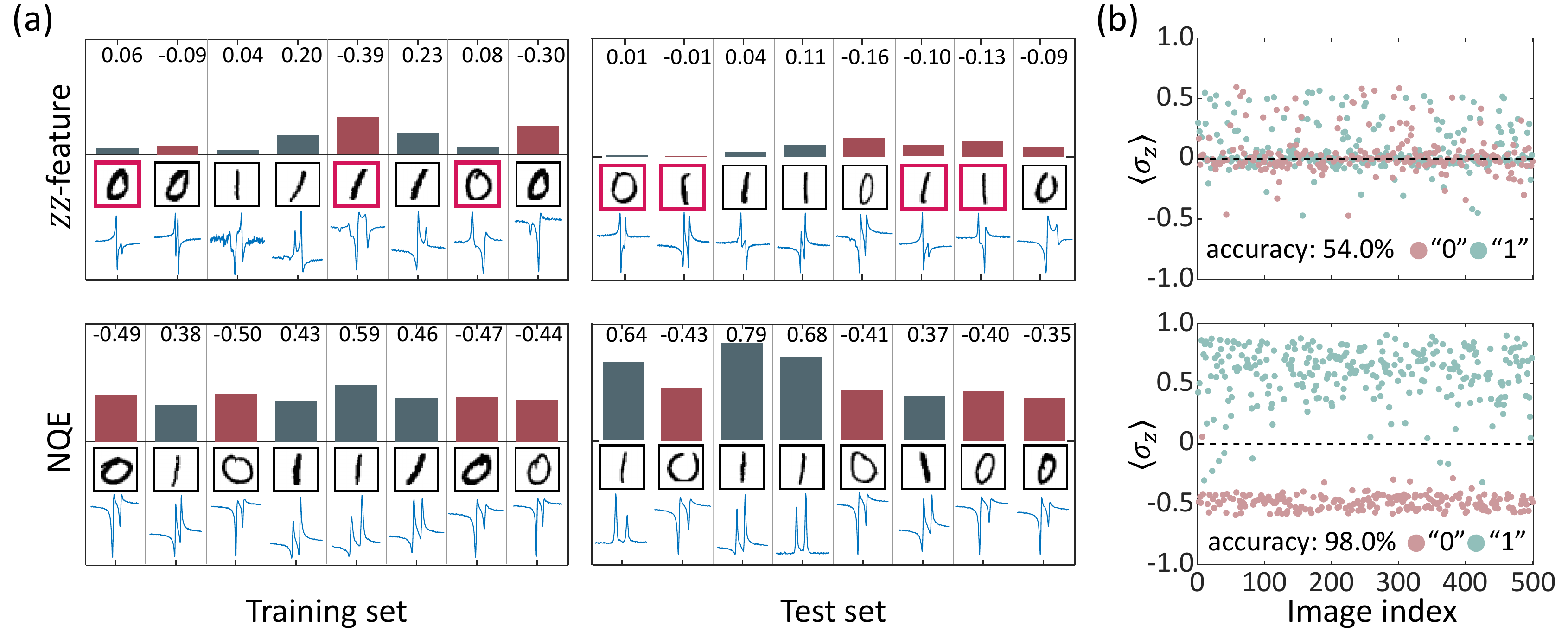}
    \caption{Experimental classification results after processing through the PQC circuit for $ZZ$-feature and NQE encoding. (a) The top and bottom panels present the results for $ZZ$-feature embedding and NQE, respectively. The numerical values correspond to the experimental $\langle\pz\rangle$, while the height of the color bars serves as a visual representation (red for negative values and green for positive). Additionally, NMR peaks provide visual indicators, with downward peaks representing ``0'' and upward peaks representing ``1''. Handwritten images outlined in red indicate incorrect classifications. These results clearly highlight the substantial improvement in classification accuracy achieved through NQE encoding. (b) Classification results for all 500 images, with the dashed line representing the decision boundary. The classification accuracy improves from 54.0\% without NQE (comparable to random guessing) to 98.0\% with NQE encoding. }
    \label{classification}
\end{figure*}

\textit{NQE-DQC1 training results}---In the experiment, we perform 15 iterations of NQE training using the DQC1 circuit.
The training loss $L_\text{NQE}$ decreases with each iteration, as shown in Fig.~\ref{nqe}(b). The triangle markers indicate the experimental result from the NMR device, while the solid line represents numberical simulations. We observe that $L_\text{NQE}$ converges close to zero around the 10th iteration.
Furthermore, the experimental results closely match the simulation results, demonstrating the high control accuracy of our experiment.

To evaluate the NQE performance, we examine how the trace distance varies over training iterations. Twenty pairs of images, each from the two categories, are randomly selected from both the training and test datasets. Ideally, we expect the trace distance for each pair to approach 1 after the NQE optimization.

We compute the trace distance for the 20 pairs at each iteration. In Fig.~\ref{nqe}(c), the solid circles represent the mean trace distance across the 20 pairs, while the error bars indicate the standard deviation. For both the training and test datasets, the gradual increase in trace distance through successive iterations shows that NQE significantly improves the separability of the embedded data. 
This substantial improvement confirms the effectiveness of NQE-DQC1 in enhancing the distinguishability between categories in classification tasks.

\textit{Classification results}---We further substantiate the effectiveness of NQE in QML by training a parameterized quantum circuit (PQC) and implementing classification tasks.
The sturcture of the PQC is shown in Fig.~\ref{PQC}(a) and consists of two parts. 
The first part embeds classical data into quantum states, while the second part trains the PQC to efficiently classify the data.
For the task of classifying handwritten digits ``0'' and ``1'', we apply a double-layer PQC with four parameters of $\theta$ and two CNOT gates, and measure $\langle \pz \rangle$ on the third qubit. The training of this PQC is carried out using the standard parameter-shift rule~\cite{Jun2017,lu2017enhancing,feng2018gradient,lin2023online}, and further experimental details can be found in Supplemental Material~\cite{supp}. .

We compare the classification performance of the traditional $ZZ$-feature embedding (without NQE)~\cite{Havlicek2019} and our protocol. This performance is visualized by measuring the loss function of the PQC, defined as $L_\text{PQC}=\frac{1}{m}\sum_i\frac{1}{2}(1-f_i\times y_i)$, at each iteration. Here, $f$ is the experimental output $\langle \pz \rangle$ on the third qubit, and $y$ is the label of the image ($y=-1$ for digit ``0'' and $y=+1$ for digit ``1''). We randomly select a batch of $m = 10$ images at each iteration to compute $L_\text{PQC}$.

The experimental results are shown in Fig.~\ref{PQC}(b). The blue and red curves represent the loss histories of $L_\text{PQC}$ from numerical simulations with and without NQE, respectively. The solid circles represent the experimental results from the NMR device. We observe that $L_\text{PQC}$ decreases significantly through iterative optimization when NQE is utilized. In contrast, the PQC training using the traditional $ZZ$-feature embedding without NQE shows only minimal improvement, highlighting the effectiveness of our NQE protocol.

Moreover, by leveraging the neural network  obtained from the NQE training, we can apply it to subsequent classification tasks on other physical systems. This highlights an additional advantage of the NQE-DQC1 protocol: its extendability. DQC1 is specifically designed for ensemble systems, making platforms such as NMR well-suited for its implementation. However, once the neural network is trained, the subsequent tasks can also be executed on other quantum platforms.
As a comparison, we train the PQC using IBM cloud superconducting processors, and the corresponding results are depicted in Fig.~\ref{PQC}(b). The red and blue stars represent IBM experimental results without and with NQE encoding, respectively. The experimental trends align well with numerical simulations, although they perform slightly worse than the NMR experiments. This further demonstrates the extendability of the NQE-DQC1 protocol in classification tasks.

After training the PQC, we evaluate its prediction accuracies. Eight images are randomly selected from both the training and test datasets, with each image encoded into quantum states via NQE before being processed through the trained PQC circuit. The expectation values $\langle\pz\rangle$ of the third spin are measured, and these values are used to determine the classification results, as shown in Fig.~\ref{classification}(a). Ideally, a negative $\langle\pz\rangle$ (red bar), corresponding to a downward NMR peak, indicates the digit ``0,'' while a positive $\langle\pz\rangle$ (green bar) or an upward peak indicates the digit ``1.'' For both the training and test datasets, the classification accuracy is significantly improved through NQE encoding, yielding more precise $\langle\pz\rangle$ values and more distinct experimental spectral features. Without NQE encoding, classification errors are evident, such as misclassifying the fifth image in the training dataset and the seventh image in the test dataset.
To further assess performance, the classification results for all 500 images in the dataset are presented in Fig.~\ref{classification}(b). With NQE, the classification accuracy achieved 98.0\%, correctly identifying 490 out of 500 images, demonstrating a remarkable improvement over the traditional $ZZ$-feature embedding. The traditional $ZZ$-feature embedding achieved an accuracy of only 54.0\%, correctly classifying just 270 images, which is comparable to random guessing. Additional experimental results on other datasets are provided in the Supplemental Material~\cite{supp} to further validate our findings.

\textit{Conclusions}---We propose the NQE-DQC1 approach for efficient classical data embedding into quantum registers. The NQE technique maximizes the trace distance between quantum states corresponding to different categories of classical data, while DQC1 enables efficient training using ensemble quantum systems, such as NMR. The protocol is validated using an NMR quantum processor, demonstrating significantly improved data embedding of handwritten images compared to the traditional $ZZ$-feature mapping approach. After training the NQE, we further implement classification tasks, achieving 98.0\% classification accuracy, in contrast to the 54.0\% accuracy obtained using traditional embedding. Additionally, we show that once the NQE is trained, subsequent QML tasks can be implemented on other physical systems, such as the IBM cloud service with superconducting circuits. Our work demonstrates that ensemble quantum systems are a powerful tool for effective QML on classical data.

\textit{Acknowledgments}---This work is supported by Pearl River Talent Recruitment Program (2019QN01X298), Guangdong Provincial Quantum Science Strategic Initiative (GDZX2303001, GDZX2200001, GDZX2403004), Institute of Information \& communications Technology Planning \& evaluation (IITP) grant funded by the Korea government (No. 2019-0-00003, Research and Development of Core Technologies for Programming, Running, Implementing and Validating of Fault-Tolerant Quantum Computing System), the Yonsei University Research Fund of 2024 (2024-22-0147), and the National Research Foundation of Korea (2023M3K5A1094813).

%

\appendix

\onecolumngrid
\newpage

\begin{center}
    \textbf{\large Supplemental Material: Neural quantum embedding via deterministic quantum computation with one qubit}\\
    \vspace{2ex}
    \text{Hongfeng Liu$^{1}$, Tak Hur$^{2}$, Shitao Zhang$^{3}$, Liangyu Che$^{1}$, Xinyue Long$^{4}$, Xiangyu Wang$^{1}$, Keyi Huang$^{1}$,} \\
    \text{Yu-ang Fan$^{1}$, Yuxuan Zheng$^{1}$, Yufang Feng$^{1}$, Xinfang Nie$^{1,4}$, Daniel K. Park$^{2,5}$, Dawei Lu$^{1,3,4}$} \\
    \vspace{1ex}
    \textit{$^1$ Department of Physics and Guangdong Basic Research Center of Excellence for Quantum Science, Southern University of Science and Technology, Shenzhen 518055, China} \\
    \textit{$^2$ Department of Statistics and Data Science, Yonsei University, Seoul 03722, Republic of Korea} \\
    \textit{$^3$ Shenzhen Institute for Quantum Science and Engineering, Southern University of Science and Technology, Shenzhen 518055, China} \\
    \textit{$^4$ Quantum Science Center of Guangdong-HongKong-Macao Greater Bay Area, Shenzhen 518045, China} \\
    \textit{$^5$ Department of Applied Statistics, Yonsei University, Seoul 03722, Republic of Korea} \\
\end{center}

\section{Deterministic Quantum Computation with One Pure Bit}
Deterministic quantum computation with one pure bit (DQC1) is a quantum computing model designed to operate with minimal quantum resources~\cite{DQC1PhysRevLett.81.5672}. In this model, the quantum system is initialized with one qubit in a pure state while the remaining $n$ qubits are in a mixed state, making it a resource-efficient framework for specific quantum computations. DQC1 can be viewed as an extension of deterministic quantum computation with pure states (DQCp), which assumes all qubits are initialized in pure states. In standard quantum computation, the final result is typically obtained through direct measurement of the first qubit. In contrast, DQCp replaces this with a process that yields the noisy expectation value of $\sigma_z^1$ for the final state. If the quantum system’s state is $\rho$, this process produces a value sampled from a distribution with mean $\langle\sigma_z^1\rangle=\text{Tr}(\sigma_z^1\rho)$ and variance $s$, where $s$ is independent of the system size. By repeating the computation and measurement process, independent samples can be collected to estimate the mean within $\epsilon$ accuracy with an error probability of at most $p$, using $O(\log(1/p)/\epsilon^2)$ repetitions.

Both DQC1 and DQCp allow the estimation of quantities involving unitary operators. For instance, if $U$ is a unitary operator implemented via a quantum circuit, DQCp enables the evaluation of $\text{Tr}(\sigma_z^1 U\ket{0} \bra{0}U^\dagger)$, and expressions like $\text{Tr}(\sigma_b U\ket{0} \bra{0}U^\dagger)$ can be computed by applying individual bit rotations during pre- and postprocessing. Similarly, in DQC1, the deviation of the initial quantum state is described by $\sigma_z^1$, and the final result is obtained through a bounded variance process, producing $\langle\sigma_z^1\rangle$. While DQCp relies on all qubits being in pure states, DQC1 achieves meaningful quantum computations with only one pure-state qubit, with the others in mixed states. This difference significantly reduces the pure-state resource requirements while preserving computational capabilities.

The computational power of both models lies in their ability to estimate specific quantities. For DQC1, this includes evaluating expressions such as $\text{Tr}(\sigma_a U \sigma_b U^\dagger)$ for any $a$ and $b$, which can be achieved by combining the model with pre- and postprocessing techniques from DQCp. These operations have a linear resource cost in the number of qubits, allowing DQC1 to maintain efficiency and practicality. As a result, DQC1 provides a compelling framework for performing quantum computations with minimal quantum resources, highlighting its potential in resource-constrained quantum systems.

\section{Limitations in Loss Optimization for Quantum Binary Classification}
In supervised learning, the primary goal is to find a prediction function $f$ that minimizes the true (expected) risk $R(f) = \mathbb{E}[l(f(X),Y)]$ with respect to a given loss function $l$, where $X$ and $Y$ are drawn from an unknown distribution $D$. Given $N$ sample data points $\lbrace (x_{i}, y_{i})\rbrace$, learning algorithms aim to identify the optimal function $f^{}$ that minimizes the empirical risk $R_N(f) = (1/N) \sum_{i=1}^{N} l(f(x_{i}), y_{i})$ within a predefined function class $F$, such that $f^{} = \arg\min_{f \in F} R_N(f)$. Quantum supervised learning algorithms leverage quantum devices to efficiently identify these prediction functions, often achieving improved performance. A common approach involves quantum neural networks (QNNs), where classical data $x$ is first encoded into a quantum state $\ket{x} = \Phi(x)\ket{0}^{\otimes n}$ via a quantum embedding circuit. A parameterized unitary operator $U(\theta)$ is then applied to transform the quantum state, followed by measurement with an observable $O$. The resulting measurement outcome serves as the prediction function $f(x; \theta) = \bra{x} U^{\dagger}(\theta) O U(\theta) \ket{x}$. Using gradient descent or similar optimization methods, the optimal parameter $\theta^{}$ is determined to minimize the empirical risk. In binary classification tasks with input $x \in \mathbb{R}^{m}$ and labels $y \in {-1, 1}$, the label of a new data point $x_\text{new}$ is predicted using the decision rule $y_\text{new} = \text{sign}[f(x_\text{new}; \theta^{})]$.

\begin{figure}[htbp]
\centering
\includegraphics[width=1\linewidth]{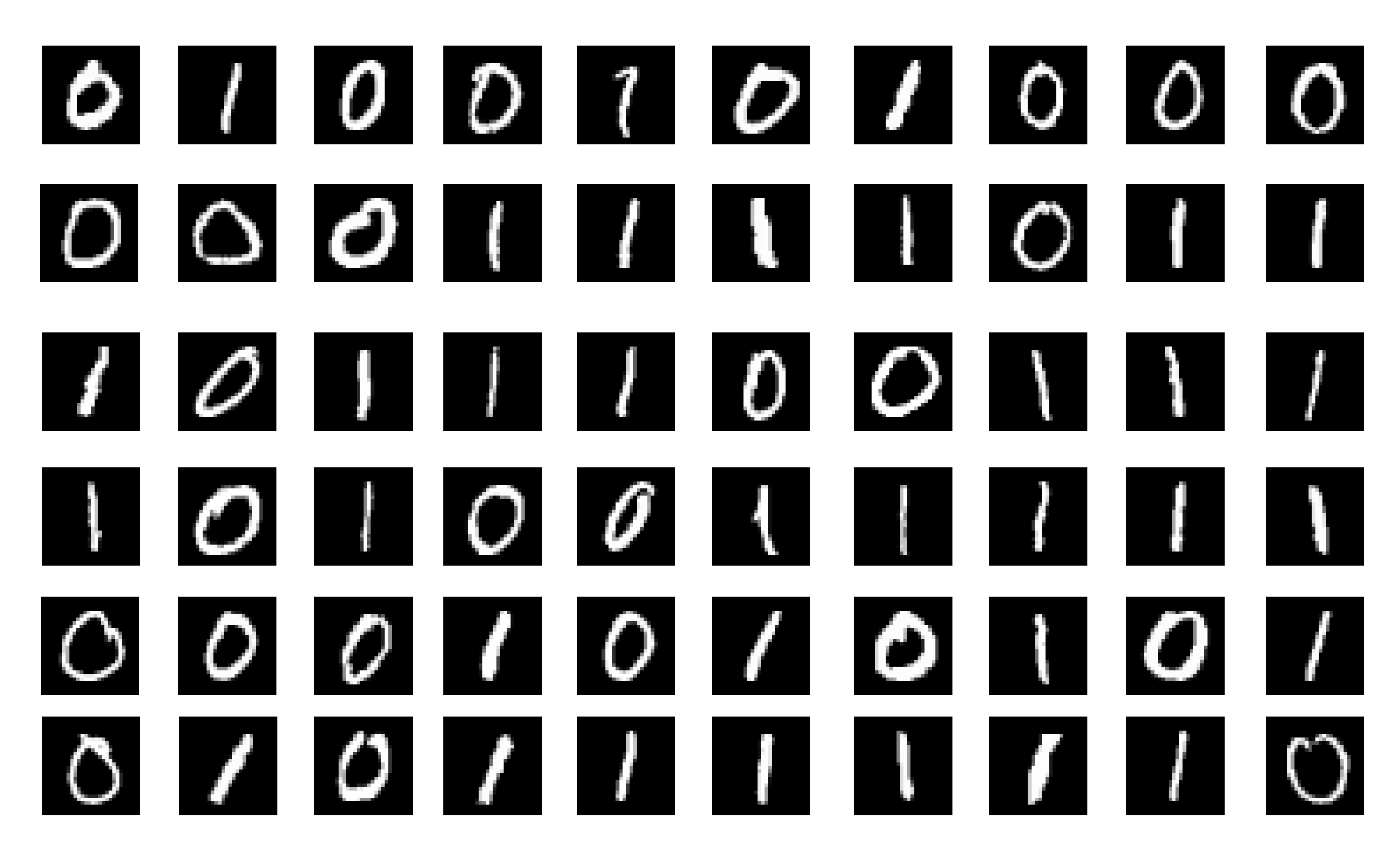}
\caption{Handwritten images of ``0'' and ``1'' from the MNIST dataset that are used for our classification experiments.}
\label{01images}
\end{figure}

This process can also be viewed as a quantum state discrimination problem, wherein two parameterized positive operator-valued measures (POVMs) $E_{\pm}(\theta) = (I \pm U^{\dagger}(\theta) O U(\theta))/2$ are defined. The probabilities of measurement outcomes $\pm 1$ for a given input $x$ are computed as $P(E_{\pm}(\theta) \vert x) = \bra{x}E_{\pm}(\theta)\ket{x}$. The decision rule then becomes $y_\text{new} = \text{sign}[P(E_{+}(\theta) \vert x_\text{new}) - P(E_{-}(\theta) \vert x_\text{new})]$. For this scenario, the probability of misclassification $l(f(x; \theta), y) = P(E_{\neg y}(\theta) \vert x)$ naturally emerges as a suitable loss function. Given a dataset $S$ consisting of $N^{-}$ samples with label $-1$ and $N^{+}$ samples with label $1$, the empirical risk is expressed as:
\begin{align} 
\label{eq:loss}
L_s &= \frac{1}{N} \bigg{[}\sum_{i=1}^{N^{-}}P(E_{+}(\theta) \vert x^{-}{i}) + \sum{i=1}^{N^{+}}P(E_{-}(\theta) \vert x^{+}{i}) \bigg{]} \nonumber \\
&\geq \frac{1}{2} - D{\text{tr}}(p^{-}\rho^-, p^{+}\rho^+), 
\end{align} 
where $ \rho^{\pm} = \sum \ket{x_i^{\pm}}\bra{x_i^{\pm}}/N^{\pm}$, $p^{\pm} = N^{\pm}/N$, and $D_{\text{tr}}(\cdot,\cdot)$ denotes the trace distance~\cite{Bae_2015}. The trace distance satisfies the contractive property for any positive and trace-preserving (PTP) map $\Lambda$:
\begin{equation} \label{eq:contractive} D_{\text{tr}}(\Lambda(\rho_0),\Lambda(\rho_1)) \leq D_{\text{tr}}(\rho_0,\rho_1), \end{equation} as shown in Ref.~\cite{e23050625}. Importantly, the lower bound of the empirical risk is determined by the trace distance between two data ensembles $p^{-}\rho^-$ and $p^{+}\rho^+$, which depends solely on the quantum embedding circuit and not on the structure of the parameterized unitary gates $U(\theta)$. The minimum loss is achieved when the POVMs $\lbrace E_{-}(\theta), E_{+}(\theta)\rbrace$ form a Helstrom measurement, making the training of a QNN effectively equivalent to finding the optimal Helstrom measurement for discriminating between the two data ensembles.

Designing a quantum embedding that maximizes the trace distance is crucial for minimizing the empirical risk, particularly in the context of noisy intermediate-scale quantum (NISQ) devices, where noise inherently reduces the trace distance between quantum states~\cite{Nielsen:2011:QCQ:1972505,wilde_2013}. Existing methods, such as combining parameterized quantum gates with conventional quantum embedding circuits, aim to create a trainable unitary embedding~\cite{lloyd2020quantum,glick2021covariant,PhysRevA.106.042431}. However, these approaches face significant challenges: they increase circuit depth, amplify noise susceptibility, and risk encountering barren plateaus that hinder scalability~\cite{10.1038/s41467-018-07090-4,thanasilp2022exponential}. Furthermore, such embeddings are fundamentally limited in their ability to enhance the maximum trace distance between quantum states. Therefore, none of the current quantum embeddings can reliably ensure effective separation of data ensembles in the Hilbert space with a sufficiently large trace distance.

\section{MNIST Dataset}
The Modified National Institute of Standards and Technology (MNIST) dataset is a widely recognized benchmark in machine learning, especially for image classification and computer vision tasks. It comprises a large collection of grayscale images of handwritten digits ranging from 0 to 9, with each image labeled according to its corresponding digit. This labeling makes the MNIST dataset ideal for supervised learning applications. The digits are written by various individuals, introducing significant variability in handwriting styles. This diversity makes the dataset an excellent tool for assessing the generalization capabilities of machine learning algorithms.

In many studies, subsets of the MNIST dataset are utilized to target specific classification tasks. For example, selecting only images of digits ``0'' and ``1'' creates a binary classification problem, simplifying the analysis and emphasizing the effectiveness of specialized algorithms. In this work, handwritten images of ``0'' and ``1'' are used for the classification experiments, with the test dataset visualized in Fig.~\ref{01images}.

\section{Experimental Details}\label{app:exp}
Our experiments were conducted using a nuclear magnetic resonance (NMR) quantum processor, which utilizes the nuclear spins within a molecule to encode qubits. Initially, we provide a detailed characterization of the NMR system, covering aspects such as sample preparation, control mechanisms, and measurement techniques. Following this, we describe the procedure for generating a pseudo-pure state (PPS).

\begin{figure}[htbp]
\centering
\includegraphics[width=1\linewidth]{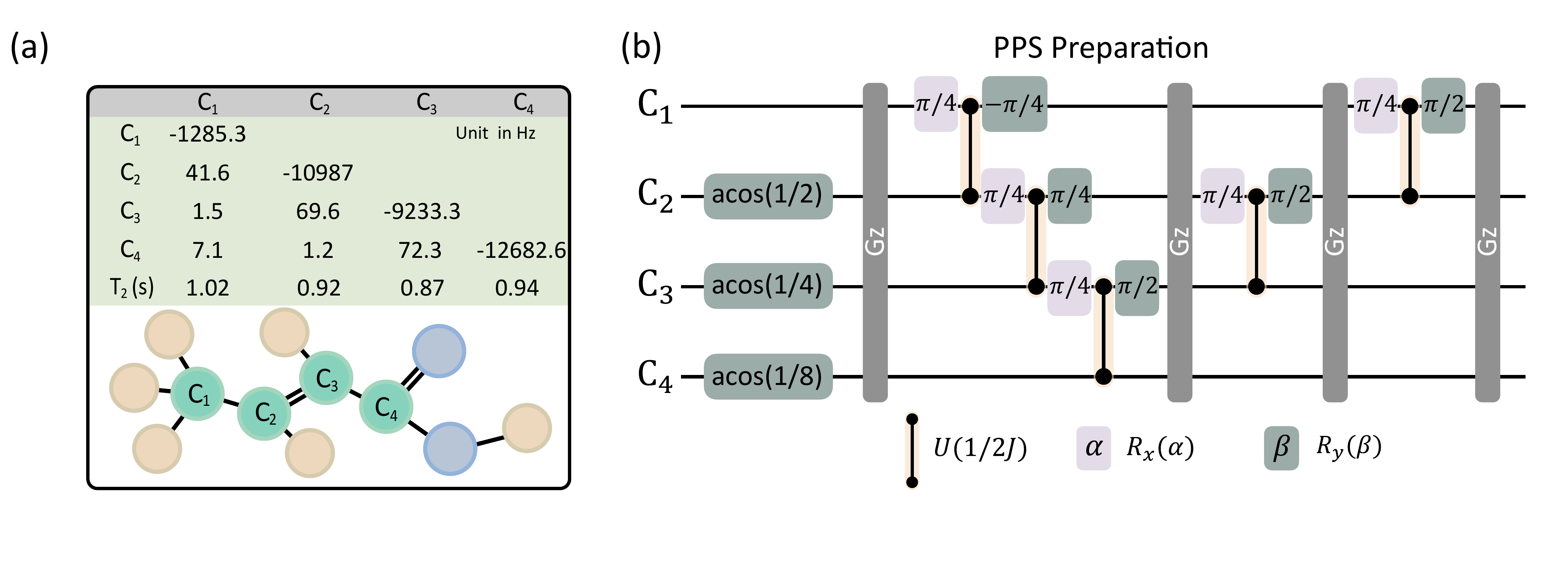}
\caption{(a) Molecular structure and key parameters of $^{13}$C-labeled trans-crotonic acid. The table lists chemical shifts (diagonal values) and scalar coupling constants (off-diagonal values, in Hz). The relaxation times $T_2$ (in seconds) are provided at the bottom of the table.
(b) NMR pulse sequence for preparing the 4-qubit system in the pseudo-pure state (PPS). The purple and green rectangles represent the $R_x$ and $R_y$ rotation gates, respectively, while the gray rectangles denote the $z$-gradient pulse. This gradient pulse is applied to eliminate all coherences from the instantaneous state.}
\label{molecular structrue&pps}
\end{figure}

\textbf{Characterization.}---In this experiment, the four-qubit quantum processor is realized using $^{13}$C-labeled trans-crotonic acid dissolved in $d_6$-acetone~\cite{nie2022experimental,park2016simulation,nie2024self,zhang2022identifying}. The molecular structure and corresponding parameters are illustrated in Fig.~\ref{molecular structrue&pps}(a). Qubits Q1 through Q4 correspond to $^{13}$C$_1$ through $^{13}$C$_4$, respectively, while the methyl group M and all hydrogen atoms were decoupled throughout the experiments. The total Hamiltonian $\mathcal{H}_\text{tot}$ for this system comprises the internal Hamiltonian $\mathcal{H}_\text{int}$ and the control Hamiltonian $\mathcal{H}_\text{con}$, expressed as: \begin{align}
{\mathcal{H}_{{\rm{tot}}}} ={\mathcal{H}_{{\rm{int}}}} +{\mathcal{H}_{{\rm{con}}}} 
 =\sum\limits_{i = 1}^4 {\pi {\nu_i}\sigma _z^i} + \sum\limits_{1 \le i < j \le 4}^4 {\frac{\pi }{2}{J_{ij}}\sigma _z^i} \sigma _z^j
 -B_1\sum\limits_{i = 1}^4 \gamma_i[\cos(\omega_{rf}t+ \phi)\sigma_x^i+\sin(\omega_{rf}t+ \phi)\sigma_y^i],
\end{align}
where $\nu_i$ represents the chemical shift of the $i$th spin, and $J_{ij}$ is the scalar coupling constant between the $i$th and $j$th nuclei. The parameters $B_1$, $\omega_{rf}$, and $\phi$ refer to the amplitude, frequency, and phase of the control pulse, respectively.

\textbf{Pseudo-pure state preparation.}---At room temperature, the thermal equilibrium state of the four-qubit NMR system is a highly mixed state, described by
\begin{equation}
\rho_{\text{eq}} = \frac{I}{16} + \epsilon\sum^4_{i=1}\sigma_z^i,
\label{eqstate}
\end{equation}
where $I$ is the $16 \times 16$ identity matrix, and $\epsilon$, representing polarization, is approximately $10^{-5}$. This mixed state is unsuitable as the initial state for quantum computation.
To address this issue, various initialization techniques, such as the spatial averaging method, line-selective transition method, time-averaging method, and cat-state method, can be employed. In our experiments, we adopted the spatial averaging method to initialize the NMR system, using the pulse sequence depicted in Fig.~\ref{molecular structrue&pps}(b). In the circuit diagram, the colored rectangles represent single-qubit rotations achieved using radio-frequency pulses, while two-qubit gates are realized through scalar coupling between spins combined with shaped pulses.

This pulse sequence transforms the equilibrium state described in Eq.~(\ref{eqstate}) into the PPS, expressed as
\begin{equation}
\rho_{\text{PPS}} = \frac{1-\epsilon'}{16}{I} + \epsilon'\ket{0000}\bra{0000}.
\label{eq:PPS}
\end{equation}
The dominant component, the identity matrix $I$, remains invariant under any unitary transformation and is undetectable in NMR experiments. This property allows the quantum system to be effectively treated as the pure state $\ket{0000}\bra{0000}$, despite its actual mixed nature.
In our experimental setup, each segment of the quantum circuit, separated by four gradient pulses, was combined into a single unitary operation. The corresponding radio-frequency pulses were designed using an optimal-control algorithm. The shaped pulses employed in the experiments had durations of 3 ms, 20 ms, 15 ms, and 15 ms, respectively, with all pulses exhibiting fidelities exceeding 99.5$\%$.

\textbf{Measurement.}---In an NMR quantum processor, the experimental sample comprises a large number of identical molecules rather than a single molecule. As a result, the measurements performed by the NMR system represent ensemble averages over all these molecules. After the quantum operation, the nuclear spins precess around the $B_0$ direction and gradually relax back to thermal equilibrium. During this precession, the nuclear spins induce an electrical signal in the $x$-$y$ plane, enabling the NMR system to measure only the transverse magnetization components, specifically the expectation values of $\sigma_x$ and $\sigma_y$. 

In this four-qubit NMR quantum processor, the signal from each nuclear spin is typically split into eight distinct peaks due to the couplings between different nuclei. Based on the principles of spin dynamics in NMR, the signal associated with each peak comprises both real and imaginary components. These components encode the expectation values of the Pauli matrices $\sigma_x$ and $\sigma_y$, respectively, for the observed spin. Consequently, the NMR system is capable of measuring the expectation values of single-quantum coherence operators, which involve $\sigma_x$ or $\sigma_y$ for the target qubit and $\sigma_z$ or $I$ for the remaining qubits.
In our protocol, the primary focus is on measuring longitudinal magnetization observables, such as $\sigma_zIII$. To achieve this, readout pulses are applied to transform these longitudinal observables into their transverse counterparts. For instance, the readout pulse $R_y^1(\pi/2)$ is used to measure $\sigma_zIII$ by converting it into $\sigma_xIII$ for detection.

\textbf{Additional classification results.}---
In addition to the recognition results discussed in the main text, we randomly selected 30 images from the test dataset to further evaluate the classification performance. In the experiments, these image data were first embedded into quantum states and subsequently processed through the trained parameterized quantum circuit. The expectation value $\langle\pz\rangle$ of the C$_3$ qubit was then measured, and these values were used to determine the classification outcomes.

\begin{figure}[htbp]
\centering
\includegraphics[width=1\linewidth]{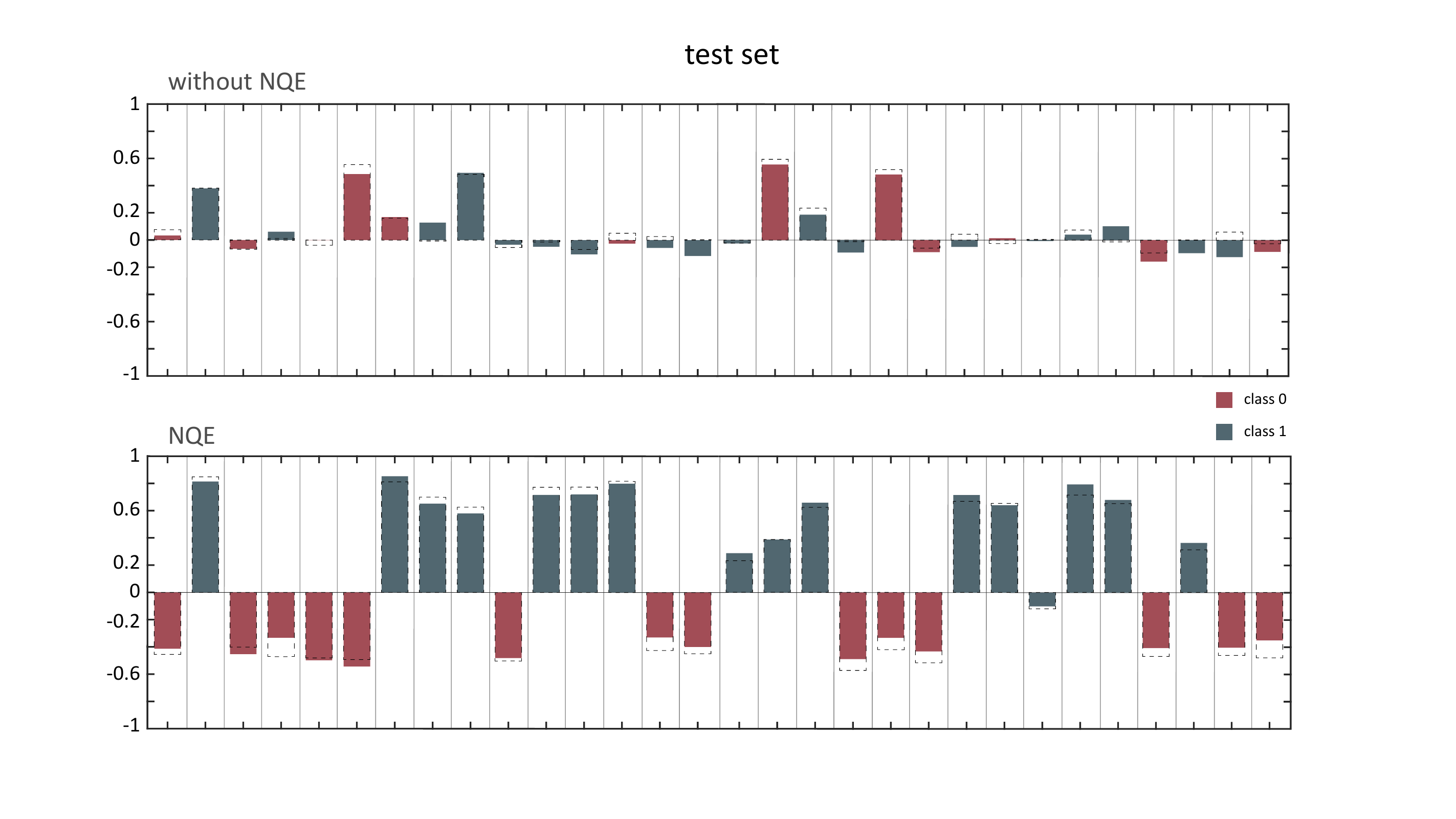}
\caption{The top and bottom panels present the results for $ZZ$-feature embedding and NQE, respectively. The numerical values correspond to the experimental $\langle\pz\rangle$, while the height of the color bars serves as a visual representation (red for negative values and green for positive).}
\label{more QCNN results}
\end{figure}

The results for both the $ZZ$-feature and NQE approaches are shown in Fig.~\ref{more QCNN results}. For class 0 and class 1, the correct expectation values are expected to be negative and positive, respectively. It is evident that the results obtained without employing NQE exhibit significant errors. For instance, some class 0 samples yield positive expectation values, while others have values so small that they become indistinguishable, highlighting the limitations of the traditional approach.

\section{additional dataset}
\begin{figure}[htbp]
\centering
\includegraphics[width=1\linewidth]{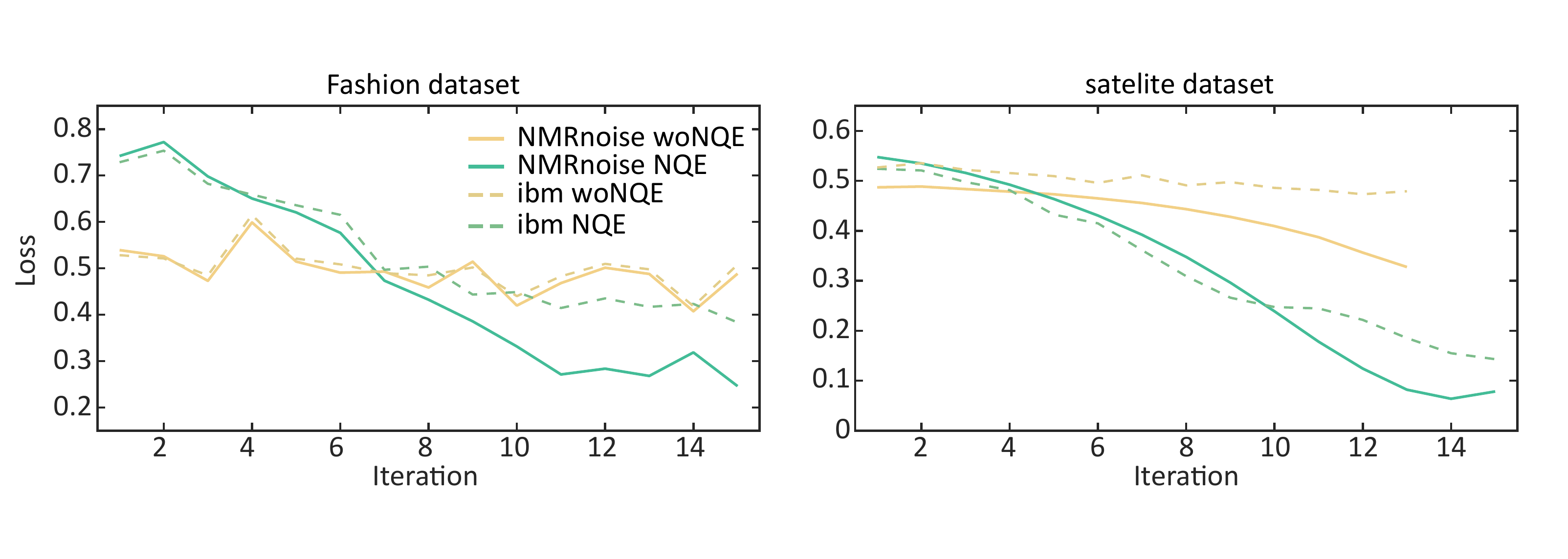}
\caption{Optimization results of the parameterized quantum circuit for the Fashion-MNIST (left) and satellite image (right) datasets. The optimization results for the parameterized quantum circuits are presented, where solid lines, dots, and dashed lines correspond to the loss $L_{\text{PQC}}$ from numerical simulations, NMR  numerical experiments, and IBM experiments, respectively. Yellow and green denote results obtained under traditional $ZZ$-feature embedding (without NQE) and with NQE encoding, respectively.}
\label{add_data_qcnnloss}
\end{figure}

We further selected Fashion-MNIST~\cite{xiao2017/online} and satellite image datasets to validate the effectiveness of our NQE protocol. 
Initially, we conducted NQE training numerically to optimize the neural network parameters $w$, incorporating NMR noise conditions into the simulations. 
Once the neural network was trained, we proceeded to train the parameterized quantum circuit.
The parameterized quantum circuit training was conducted on a 127-qubit IBM Eagle processor (\textit{ibm\_yonsei}) and through numerical simulations incorporating NMR noise. The training of the parameterized quantum circuit and its evaluation on test datasets were conducted both with and without NQE.
The results are presented in the Fig.~\ref{add_data_qcnnloss}.

\end{document}